\documentclass[10pt, sigconf]{acmart}

\usepackage[linesnumbered,ruled,vlined]{algorithm2e}
\usepackage{algpseudocode}
\usepackage{amsthm}
\usepackage{amsmath}
\usepackage{balance}  
\usepackage{booktabs}
\usepackage{bold-extra}
\usepackage{enumitem}
\usepackage{graphicx}
\usepackage{hyperref}
\usepackage{mathtools}
\usepackage[linewidth=1pt]{mdframed}
\usepackage{multirow}
\usepackage{pifont}
\usepackage{colortbl}
\usepackage{stmaryrd}
\usepackage{subcaption}
\usepackage{tabulary}
\usepackage{tabularx}
\usepackage[utf8]{inputenc}

\SetAlFnt{\small}
\SetAlCapFnt{\small}
\SetAlCapNameFnt{\small}
\SetAlCapHSkip{0pt}
\IncMargin{-\parindent}

\newcommand{\ainv}{$\alpha$-investing }

\newcommand*{\smparagraph}[1]{\vspace{0.3cm}\noindent{\bf #1.}}

\newcommand{\new}[1]{{\em #1\/}}

\theoremstyle{definition}

\numberwithin{remark}{section}
\numberwithin{theorem}{section}
\numberwithin{definition}{section}

\definecolor{colorlow}{HTML}{00b894} 
\definecolor{colorhigh}{HTML}{FC427B} 
\definecolor{bluegray}{HTML}{dcdde1} 

\newcommand*\nocolor[1]{\cellcolor{white!0}}

\newcommand*\name{\textsf{STAR}\xspace} 
\newcommand*\server{\mathcal{S}} 
\newcommand{\serverset}{\{\server_1, \dots, \server_\np\}}

\newcommand{\fpparam}{f} 
\newcommand{\ssparam}{\kappa} 
\newcommand{\Ds}{\mathcal{D}} 

\newcommand*\np{k} 
\newcommand*\nrow{n} 
\newcommand*\ncat{m} 
\newcommand{\Log}{\mathsf{TestLog}}
\newcommand{\code}{\mathcal{T}}
\newcommand{\codeid}{\sigma}
\newcommand{\cert}{\psi}
\newcommand{\result}{\tau}
\newcommand{\pval}{\mathsf{p}}
\newcommand{\logid}{\rho}

\newcommand*\lsetup{\mathcal{L}_\mathsf{s}} 
\newcommand*\lcomp{\mathcal{L}_\mathsf{c}} 

\newcommand{\enc}[1]{[#1]} 
\newcommand{\denc}[2]{[#1_{#2}]} 

\newcommand*\ttest{Student's t-test\xspace}					
\newcommand*\pearson{Pearson's correlation\xspace}	
\newcommand*\chisq{Chi-squared\xspace}
\newcommand*\ftest{ANOVA F-test\xspace}

\usepackage{booktabs} 
\usepackage[htt]{hyphenat}

\renewcommand\footnotetextcopyrightpermission[1]{} 
\setcopyright{none}
\begin{document}


\title{STAR: Statistical Tests with Auditable Results\\  \LARGE{System for Tamper-proof Hypothesis Testing}}

\author{Sacha Servan-Schreiber}
\affiliation{%
	\institution{MIT CSAIL}}
\author{Olga Ohrimenko}
\affiliation{%
	\institution{Microsoft Research}}
\author{Tim Kraska}
\affiliation{%
	\institution{MIT CSAIL}}
\author{Emanuel Zgraggen}
\affiliation{%
	\institution{MIT CSAIL}}
\renewcommand\shortauthors{Servan-Schreiber, S. et al}


\begin{abstract}

\noindent We present \name: a novel system aimed at solving the complex issue of ``p-hacking'' and false discoveries in scientific studies.  \name provides a concrete way for ensuring the application of false discovery control procedures in hypothesis testing, using mathematically provable guarantees, with the goal of reducing the risk of data dredging. \name generates an efficiently auditable certificate which attests to the validity of each statistical test performed on a dataset. \name achieves this by using several cryptographic techniques which are combined specifically for this purpose. Under-the-hood, \name uses a decentralized set of authorities (e.g., research institutions), secure computation techniques, and an append-only ledger which together enable auditing of scientific claims by 3rd parties and matches real world trust assumptions. We implement and evaluate a construction of \name using the Microsoft SEAL encryption library and SPDZ multi-party computation protocol. Our experimental evaluation demonstrates the practicality of \name in multiple real world scenarios as a system for certifying scientific discoveries in a tamper-proof way. 

\end{abstract}

\maketitle


\section{Introduction}
According to a 2016 Nature Magazine survey, over 70\% of researchers failed to reproduce published results of other scientists and \emph{over 50\% failed to reproduce their own published results}~\cite{baker20161}. 
The ``Replication Crisis'', plaguing almost all scientific domains, has serious and far reaching consequences on the continued progress of scientific discoveries. Unfortunately, solutions addressing the problem are few and often ineffective for two reasons: 1) current solutions either fail to take into account real world trust assumptions (e.g., by trusting researchers to carefully apply false discovery control protocols) or 2) are overly restrictive (e.g., by requiring independent replication of results prior to publishing or pre-registration of hypotheses). Moreover, these solutions fail to take into account modern approaches to data analysis, specifically the abundance of existing data and means by which to explore it, and impose overly stringent requirements. 

The replication crisis is, in part, a direct result of these problems and what is formally known as the Multiple Comparisons Problem (MCP). With every hypothesis tested over a dataset (using any type statistical testing procedure), there is a small probability of a chance, i.e., false positive, discovery with no real basis to the population being studied. With every additional statistical test performed on the data, the chance of encountering such a random correlation increases. 

This can be intentionally exploited to ``fabricate'' significant discoveries and, if done in systematically, is referred to as ``HARKing``~\cite{kerr1998harking}, ``p-hacking''~\cite{head2015extent} or ``data dredging''.

While a variety of statistical techniques exist to control for the MCP by setting threshold on the false discovery rate (FDR), i.e., the ratio of false-positives to true-positives over a sequence of hypotheses~\cite{dunn1961multiple,benjamini1995controlling}, there is surprisingly almost no support to ensure that researchers and analysts actually use them. 
Rather individual research groups rely on often varying data analysis guidelines and trust in their group members to abide by the control procedures correctly which, unfortunately, rarely works in the real world. This is because 1) making even one simply mistake in the application of the control procedure can result in a false discovery and 2) there is no means of guaranteeing that each researcher carefully applied the control procedure (or, didn't intentionally deviate from the procedure to get a ``significant'' (false) discovery). 
Things get even worse when the same data is analyzed by several institutions or teams since guarding against false discoveries requires a coordinated effort. 
It is currently close to impossible to reliably employ statistical procedures, such as the Bonferroni~\cite{dunn1961multiple} method, that guard against p-hacking across collaborators.
It only requires one member to ``misuse'' the data (intentionally or otherwise) and detecting, let alone recovering from such incidents is next to impossible. 
This problem is perhaps further exacerbated by the pressure on PhD students and PIs to publish \cite{neill2008publish}, ``publication bias''~\cite{dickersin1987publication} as papers with significant results are more likely to be published, and the increasing trend to share and make datasets publicly available for any researchers to use in studies.  
Therefore, after examining the state of affairs, it is perhaps not surprising that scientific community is plagued by false discoveries~\cite{begley2012drug,ioannidis2005contradicted,john2012measuring,ioannidis2005most}.

To illustrate this problem concretely, consider a publicly available dataset such as MIMIC III \cite{johnson2016mimic}. 
This dataset contains de-identified health data associated with $\approx40,000$ critical care patients.
MIMIC III has already been used in various studies \cite{mayaud2013dynamic,ghassemi2014data,henry2015targeted} and it is probably one of the most (over)analyzed clinical datasets and therefore prone to ``dataset decay''~\cite{thompson2019dataset}.
As such, any new discovery made on MIMIC runs the risk of being a false discovery.
Even if a particular group of researchers follow a proper FDR control protocol, there is \emph{no control} over happens across different groups and tracking hypotheses at a global scale poses many of its own challenges.
It is therefore hard to judge the validity of any insight derived from such a dataset~\cite{thompson2019dataset}.

A solution to guarantee validity of insights commonly used in clinical trials --- preregistration of hypotheses \cite{cockburn2018hark} --- falls short in these scenarios since the data is collected upfront \emph{without} knowing what kind of analysis will be done later on. 
Perhaps more promising is the use of a hold-out dataset. 
The MIMIC author, for example, could have released only $30$K patient records as an ``exploration'' dataset and hold back $10$K records as a ``validation'' dataset. 
The exploration dataset can then be used in arbitrary ways to find interesting hypotheses. 
However, before any publication is made by a research group using the dataset, all hypotheses must be (re)tested for statistical significance over the validation dataset. 
Unfortunately, in order to use the validation dataset more than once, we run into the same probem: every hypothesis over the validation dataset has to be tracked and controlled for. 
Furthermore, the data owner (the MIMIC author in this case) needs to provide this hypothesis validation service. 
This is both a burden for the data owner as well as a potential risk. 
Researchers need to trust the data owner to apply necessary control procedures and to objectively evaluate their hypotheses, which, unfortunately does not always align with real world incentive structures. 

The above example illustrates the motivation behind the need for a system that addresses these problems. 
{\bf With \name, our goal is to create a system that guarantees the validity of statistical test outcomes and allows readers (and/or reviewers) of publications to audit them for correctness, all without introducing unnecessary burdens on data provider and researchers.} 
Using cryptographic techniques
to certify outcomes of statistical tests and by introducing a decentralized authority, we eliminate the risk of data-dredging (intentional and otherwise) by researchers using a dataset. 
\name can be used in various settings, including cases where the data is public and only the hold-out data is fed into \name (as in the example above), in settings where a few research groups collaborate on combined data, or even within single teams where lab managers can opt to use \name as a way to prevent unintentional false discoveries, assign accountability and foster reproducibility.

\subsection{Contributions}
 
\begin{itemize}
	\item We present a novel system for preventing p-hacking using cryptographic techniques which provide (mathematical) guarantees on the validity of each tested hypotheses during analysis, even in settings where researchers are not trusted to apply control procedures correctly, while also ensuring full auditability of all results obtained through \name. 
	\item We implement and evaluate \name on four widely used statistical tests (\ttest, \pearson, \chisq and \ftest) to demonstrate the applicability of \name to real world scenarios. 
	\item We describe how a common false discovery control procedure known as \ainv can be applied with \name to provide full control over the data analysis phase in a certifiable manner. 
\end{itemize}

 To the best of our knowledge, \name is the first cryptographic solution to the problem of p-hacking by providing guarantees (in the form of tamper-proof certificates) on the validity of all insights gleaned from a dataset. We believe that \name is the first system to address the long standing problem of discovery certification across scientific domains and achieves this with minimal overhead on researchers and data providers.


\begin{figure*}[t]
	\centering
	\includegraphics[width=1.00\textwidth]{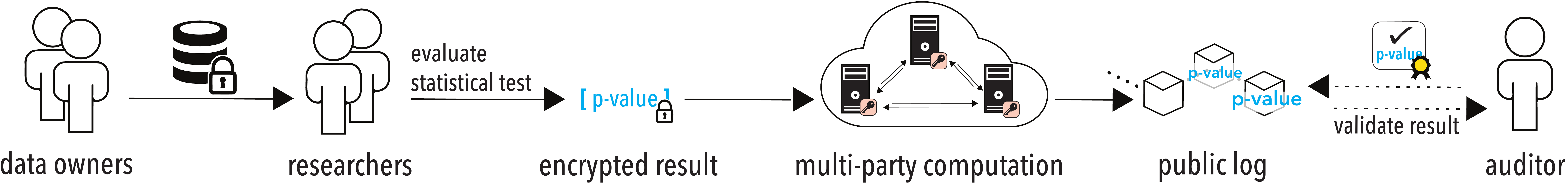}
	\caption{High-level overview of \name. Data owners upload datasets to the network of computing servers. Researchers proceed to compute statistical tests over the dataset using secure computation that is logged on a public log. An auditor can examine the log to determine the validity of a given discovery. }
	\label{fig:overview}
\end{figure*}

\section{Design}
\label{sec:design}
Our design of \name is motivated by the following observations.
The data owner cannot release a dataset~$\Ds$ to the researchers directly
since it creates a possibility for p-hacking
(e.g., researchers can run tests privately and report only
favorable results without controlling for false discoveries). Therefore, any system which addresses p-hacking must ``hide'' the raw data from researchers. With this in mind, 
we begin by describing several initial design ideas and discuss their limitations. This will serve as a motivation for why we choose the design and construction described in~\autoref{subsec:designmain} and ~\autoref{sec:protocol}. We then describe use some foreseeable use cases of our design in~\autoref{sec:scenarios}. 

\subsection{Strawman Designs}

\smparagraph{The Trusted Authority Scheme}

The simplest scheme is to assume a \emph{trusted authority} (3rd party) which has full access to the unencrypted dataset. 
Researchers then perform statistical tests on the dataset by querying the authority who executes the computations on their behalf and returns only the result of the test. 

Unfortunately, there are two immediate problems with such a design. The solution requires that the data owner, the researchers, and the auditors \emph{trust} the authority when it comes to storing the dataset, correctly executing statistical tests, and applying the control procedure. 
While this can work in certain isolated cases, often real world parties have conflicting incentives that make finding such an authority challenging. For example, consider a drug company releasing a dataset for public use but researchers don't fully trust the company to evaluate each hypothesis objectively. Moreover, there is not way to verify the actions of the authority, i.e., ensure that each statistical test was computed correctly and all p-values reported, etc., meaning that it must be trusted blindly. 

\smparagraph{The Secure Enclave Scheme}
A more viable alternative to the above scheme is to have a \emph{trusted secure processor}~\cite{subramanyan2017formal} with a strict interface which only allows for the evaluation of statistical tests. In such a scheme, the data owner encrypts the dataset using the enclave's public key which allows the enclave to execute statistical tests by first decrypting the dataset in secure memory and then evaluating the statistical test. The enclave can then reveal the result along with a digital signature authenticating the computation. However, such a design suffers from similar problems as the design involving a singe trusted authority: namely, it is now the secure enclave and the machine on which it runs that have to be trusted in the process to guarantee computation are performed~\cite{pass2017formal, choudhuri2017fairness}.

\smparagraph{The Multi-party Computation Scheme}
The secure enclave approach motivates the idea of distributing trust using multiple parties. For example, instead of having one entity in control of the data, a dataset can be shared among a set of parties which can collectively compute over their shares but cannot individually access the shared data~\cite{shamir1979share,bogdanov2007foundations,bogdanov2008sharemind}. This takes care of the guaranteed output problem and provides the means for ensuring all computations are accounted for provided some threshold number of parties follow protocol correctly. While on the surface this may appear as good solution, there are several barriers when it comes to practicality. Computing statistical tests over such a shared dataset requires \emph{general multi-party computation} which quickly becomes impractical due to high network overheads needed to evaluate functions using the secret shared data~\cite{keller2016mascot, damgaard2013practical, bogdanov2008sharemind}. In general, multi-party computation is only practical when it comes to evaluating simple functions with few shared inputs~\cite{keller2016mascot, damgaard2013practical} 
This motivates our main design which uses minimal multi-party computation and other secure computation techniques to create a practical scheme.

\subsection{\name Framework}
\label{subsec:designmain}
With the initial design attempts described above, we are ready describe the design of \name, which addresses all the problems mentioned in the strawman ideas. We desire a \emph{general} solution which fits well with real world trust assumptions and incentives. The design of \name achieve these requirements by using two cryptographic building blocks. Specifically, we use \new{secure computation} and a \new{distributed ledger} (covered in the following section), which are combined in a special way to achieve the desired goal building a system for conducting auditable hypothesis testing while remaining practical.~\autoref{fig:overview} provides a design overview of the system. 

At a high level, the data owner encrypts a dataset~$\Ds$ and releases the (encrypted) dataset publicly or to a set of researchers. The encryption scheme used is ``special'' in the sense that encrypted values can be used as inputs to functions, and the functions evaluated over the encrypted data, without knowledge of the secret key needed to decrypt the values. Such schemes are called \new{homomorphic encryption schemes}.
Researchers use the encrypted dataset, denoted~$\denc{\Ds}{}$, to compute statistical tests (encoded as an arithmetic circuit) using the homomorphic properties of the scheme.  
However, one problem remains: once the statistical test is evaluated, the result is still encrypted and researchers do not have the secret key needed to decrypt it. One trivial solution is to send the result to the data owner who then uses the secret key to decrypt the result. However, this requires the data owner to be online in order process such decryption requests and furthermore places full trust in the data owner to correctly report results, which, as we mention above we would like to avoid. Moreover, this can be especially problematic in scenarios where the dataset is comprised of multiple data sources since there is only one secret key.   

In \name, we distribute the power to decrypt ciphertexts to a set of parties (servers) which each hold a share of the secret key, but, crucially, cannot decrypt on their own using their share of the key. Parties can decrypt if and only if they ``agree'' to do so collectively; thus preventing any subset of rogue parties from decrypting results on their own and colluding with researchers. 

When parties decrypt the statistical test result, they also keep a public log of the action which guarantees that each revealed
statistic is recorded in a tamper-proof and auditable way. This latter requirement ensures that all statistical tests executed on $\denc{\Ds}{}$ are recorded, \emph{in sequence}, and makes it possible for \emph{anyone} to verify whether false discovery control procedures were applied correctly. We elaborate on these important requirements in subsequent sections.


\section{Use Cases}
\label{sec:scenarios}
We describe several distinct and highly relevant scenarios where \name is directly applicable. We hope these scenarios provides evidence for both the utility of \name and the relevance of the system for addressing false discoveries.

\smparagraph{First scenario} A data owner wishes to release a dataset for public research use but wants to prevent the data from being overanalyzed and false discoveries being extracted from the data as a result. As is usually done in such cases, the data owner partitions the dataset into an \emph{exploration} and a \emph{validation} dataset. The exploration dataset is released to the public without any restrictions on how statistical tests should be computed. The validation dataset is encrypted in such a way that results computed on the dataset are only made available through \name: researchers wishing execute tests on the validation dataset must do so by requesting the parties in \name to reveal the result. 

 Using the exploration dataset, researchers independently form hypotheses about the data. Prior to publishing the results, they validate each hypothesis over the validation dataset via \name. This guarantees that all validated hypothesis remain auditable and all computed p-values are accounted for in a tamper-proof, timestamped sequence on the log maintained by the decrypting parties. 
 The publication can then be validated by auditors (e.g., reviewers) to ensure that the researchers' hypothesis was accepted (resp. rejected) in a way that controls for false discoveries and that the insights therein are valid. This ensures that the publication's claims are statistically significant (within the alleged error bounds)  and \emph{guarantees} that no p-hacking or other forms of data dredging occurred during the analysis.
 
\smparagraph{Second scenario} \name also extends to setting where multiple data owners contribute to a collective dataset and want to ensure that their data is not misused. Consider a scenario where two separate research groups (that potentially don't trust each other with their data) have datasets consisting of similar attributes. For example, consider the Food and Drug Administration (FDA) in the United States and Chinese Food and Drug Administration (CFDA) in China.  Assume that both administrations have data about a new drug claiming to cure Alzheimer’s disease. Both the FDA and CFDA use their own respective datasets to explore the data and release their respective datasets in encrypted form such that tests can only be computed through \name. Once both organization have found interesting hypotheses in their own datasets, they use the other organization's data as a validation dataset by testing hypotheses through \name. In other words, each organization uses its own data to determine interesting insights and proceeds to validate each other's hypotheses using the other organization's data. Since all validations are logged in \name, results remain auditable. Moreover, the data of each organization remains \emph{private} (e.g., the FDA does not learn CFDA's data, except for the result of a statistical test). 

\smparagraph{Third scenario}
A data owner wishes to release sensitive data for research purposes but is required to comply with stringent regulatory demands for privacy (e.g., as required by the European GDPR regulation~\cite{voigt2017eu}). Even allowing researchers to compute statistical tests can be a violation under such privacy laws. This results in potentially useful data being left outside of the reach of the research community. Using \name, however, it becomes possible to accommodate privacy needs by only allowing researchers to test for \emph{statistical significance} (i.e., not even revealing the p-value for a statistical test, which we describe in a following section). \name can be extended to only reveal whether or not a given hypothesis should be accepted (resp. rejected) while simultaneously controlling for false discoveries.
\section{Technical Details}
We now turn to describing the security properties we require and explain how \name can be used to certify the correct application of discovery control procedures.
\subsection{Security Goals}
\label{subsec:secgoals}
In order adequately present and analyze the guarantees of \name, we must first establish a set of security goals. 
At a high level, \name aims to achieve the following goals which guarantee correctness of the system. We outline the goals in this section and formalize them in~\autoref{sec:security}.

\smparagraph{Correctness} Statistical tests computed on the encrypted dataset $\denc{\Ds}{}$ must be correct, i.e, the p-value of a test computed in \name has to be equivalent to the p-value computed through standard statistical packages (e.g., MATLAB and SciPy) over the unencrypted $\Ds$. 

\smparagraph{Confidentiality} The only information revealed about $\Ds$ is the 
	results of statistical tests which ensures that false discoveries are fully controlled for and no hypothesis can be tested outside of 
	\name's interface. In other words, \name does not ``leak'' data about the $\Ds$ when statistical tests are evaluated (more than the result of the statistical test).  

\smparagraph{Access control} 
In many cases, it is desirable to provide access to only a set of approved researchers. We require that \name have a mechanism in place for enforcing access control, though it is not an inherent requirement. If an access control policy is set, then statistical tests on $\Ds$ can only be executed by the set of approved researchers. 

\smparagraph{Auditability} For all hypotheses evaluated through \name, the resulting p-values can be checked for their statistical significance with respect to the false discovery control procedure used. {\bf Crucially, anyone can audit the correctness of a statistical test and can do so using only publicly available information and without interacting with researchers or the computing parties. }
	
\subsection{Threat Model}
\name provides provable guarantees to the four properties outlined in ~\autoref{subsec:secgoals} under the following threat model. Our security assumptions are with respect to the data owner(s), computing parties (e.g., different institutions and research labs), and researchers evaluating their hypotheses through \name.

\smparagraph{Data Owner} We make the necessary assumption that the data owner does not collude with the researchers who run the statistical tests. This assumption is required given that a malicious owner has unfettered access to the (unencrypted) dataset $\Ds$ and can thereby trivially avoid executing tests through \name, only ``testing'' results known to be significant ahead of time without applying the necessary control procedure.

\smparagraph{Computing Servers} 
Computing parties are collectively in possession of the decryption key, however, none of the parties individually have access to the unencrypted data $\Ds$. The dataset remains unknown to the servers provided at least one of the servers follows protocol correctly. If the parties are also maintaining the distributed log on which statistical test results are recorded, we assume that a majority of the parties are honest when it comes to maintaining the correctness of the log. 

\smparagraph{Researchers} 
We assume that researchers are malicious and interested in gaining as much information on $\Ds$ as possible (for the purpose of bypassing control procedures and falsely certifying discoveries). To this end, we assume that researchers are interested in abusing the system with the goal of extracting as many insights as possible and avoiding applying the false discovery control procedure. To provide accountability and prevent other forms of attacks, we require a secure public key infrastructure which identifies individual researchers by their public key and assume that all messages exchanged between researchers and computing servers are digitally signed during protocol execution.

\subsection{Controlling False Discoveries}
\label{subsec:mcpcontrol}
An important first step in realizing the construction of \name is understanding how false discovery control procedures are used in the data analysis phase. 
In order to certify discoveries, it is necessary to use the (ordered) sequence of p-values computed over dataset. The sequence is then used to bound the false discovery rate (FDR) during or after analysis.

While standard procedures such as Bonferroni~\cite{dunn1961multiple} must be applied \emph{a posteriori} of the data analysis (once all hypotheses have been tested), in \name, we desire a method for controlling the FDR in a streaming fashion, ideally without knowledge or restriction on future tests. 
Fortunately, there is a common false discovery control procedure known as \ainv~\cite{foster2008alpha} which achieves just that. The \ainv procedure is a standard choice for controlling false discoveries in practice when the total number of hypotheses is unknown ahead of time (e.g., in exploration settings)~\cite{zhao2017controlling,zhou2005streaming}. We describe the intuitive version of the procedure below and we refer the reader to~\cite{foster2008alpha,zhao2017controlling} for more formal definitions and proofs which are outside the scope to understanding this work.

The $\alpha$-investing procedure works by assigning, to each hypothesis, a ``budget'' $\alpha$ from an initial ``$\alpha$-\emph{wealth}.'' If the p-value of the null hypothesis being considered is above some $\alpha$ the null hypothesis is accepted and some budget is lost, otherwise the null is rejected and the available budget increases. Choosing $\alpha$ depends on the ``investment strategy''~\cite{zhao2017controlling} and is ultimately left as a choice for the analyst. We present a common investment strategy in Procedure~\ref{alg:investmentrule} which allows for a (theoretically) unbounded number of statistical tests to be executed over the dataset. 

More formally, for the $j$th null hypothesis, $H_j$, being tested, $H_j$ is assigned a budget $\alpha_j>0$. Let $\pval_j$ denote the p-value associated with the result of the statistical test used to determine the statistical significance of $H_j$. The null is \emph{rejected} if $\pval_j \leq \alpha_j$, otherwise it is \emph{accepted}. 
In the case that  $H_j$ is rejected, the testing procedure obtains a ``return'' on investment $\gamma \leq \alpha$. 
On the other hand, if the $H_j$ is accepted, $\alpha_j/(1-\alpha_j)$ wealth is subtracted from the remaining $\alpha$-wealth. Therefore, the remaining wealth, denoted $W(j)$, after the $j$th statistical test is set according to:

\begin{equation}
\label{eq:alpharul}
W(j+1) = \begin{cases}
W(j) + \gamma &\text{ if }p_j\leq \alpha_j,\\
W(j) -\frac{\alpha_j}{1-\alpha_j} &\text{ if }p_j> \alpha_j
\end{cases}
\end{equation}

\begin{algorithm}
	\SetAlgorithmName{Procedure}{protocol}{}
	\caption{\texttt{InvestmentRule($\alpha, \beta, \gamma$)}~\cite{zhao2017controlling}}
	\label{alg:investmentrule}
    
    $W(0) = \alpha$\tcp*{initial wealth}
                               
	\ForEach{$i \gets 1,2,\dots $}{
	 	$\alpha_i = \min\left(\alpha, \frac{W(i-1)\left(1-\beta\right)}{1 + W(i-1)\left(1-\beta\right)}\right)$\;
	 	
	 	$p_i \gets test(H_i)$\tcp*{p-value for hypothesis $H_i$}
	 	
       	\If{$p_i < \alpha_i$}{
           $W(i)= W(i-1)+\gamma$\;
           }
       
        \uElse {
           $W(i)= W(i-1) - \frac{\alpha_i}{1 - \alpha_i}$\;
        }
	}
\end{algorithm}

In \name, the parameters $W(0)$, $\alpha$, $\beta$, and $\gamma$
can be set as static constant or determined by more complex mechanisms, as seen fit. 
{\bf The important takeaway in understanding \name is that given 
all p-values computed on $\boldsymbol{\Ds}$, up to the current test, it is easy to determine whether any given hypothesis should be accepted (resp. rejected) based on the corresponding p-value in order to bound the number of false discoveries to $\boldsymbol{\alpha}$ overall.} We illustrate the process in following example.

\begin{center}
\fbox{
\begin{minipage}{0.965\columnwidth}
\paragraph{\bf Illustrative example:} Consider three hypotheses $H_1, H_2$, $H_3$, with p-values $\pval_1=0.0012, \pval_2 = 0.2331$, and $\pval_3 = 0.0049$, respectively. Set the initial alpha wealth $W(0) = 0.05$, fix $\gamma = W(0)/4$, and let $\beta = 0.25$. 
After testing hypothesis $H_1$ we obtain $\pval_1<\alpha_1$ so $H_1$ is rejected and we obtain a ``return on investment.'' Now $W(1) = W(0) + \gamma = 0.0625$. For the second test, we obtain $\pval_2 > \alpha_2$ so the $H_2$ is accepted and the remaining wealth is $W(2) = W(1) - \frac{\alpha_j}{1-\alpha_j} \approx 0.01$. For the third test, we get $\pval_3 < \alpha_j$ so $H_3$ is rejected and  $W(3) \approx 0.022$. Examining the sequence of p-values, it is trivial to verify the validity of discoveries $H_1$ and $H_3$ by reapplying Procedure~\ref{alg:investmentrule} with all three computed p-values.
\end{minipage}}
\end{center}

\subsection{Certification of Hypotheses}
The goal of \name is to certify, in a tamper-proof way, that the insights gained from the available data are valid. Taking a birds-eye view, for every test computed through \name, a publicly auditable certificate $\cert$ is produced which attests to the application of the \ainv procedure during hypothesis testing process.
Formally, we define a certificate of a statistical test $\code$ as a signed tuple $(\logid, \codeid, \result)$ where $\logid$ is the \emph{test index}  (i.e., the test's order among all the tests that have been executed so far on $\Ds$), $\codeid$ is the identifier of the statistical test function $\code$, and $\result$ is the result of executing $\code(\Ds)$, i.e., the test statistic.  Given that the certificate contains both the test index $\logid$ and resulting p-value, it is easy to verify (in conjunction with previous certificates) whether the false discovery control procedure was applied correctly by (re)applying Procedure~\ref{alg:investmentrule} and verifying the conclusion researchers came to during analysis. {\bf Note that the statistical tests do not need to be re-evaluated in order to verify the correctness of results; the p-value only has to be computed once.}


\begin{figure}[t]
	\centering
	\includegraphics[width=1.00\columnwidth]{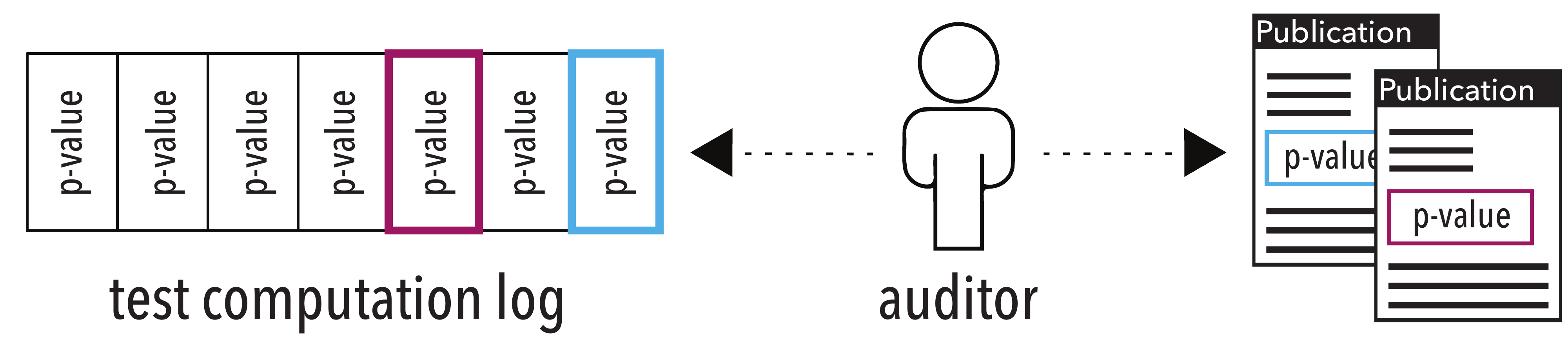}
	\caption{Auditor verifies the validity of reported p-values by examining the values stored on the log and ensuring that the \ainv procedure was applied correctly.}
	\label{fig:logaudit}
\end{figure}

\section{Construction}
\label{sec:protocol}
We are now ready to describe the system in detail. We start by describing the cryptographic tools used in realizing \name and then formalize the different components of \name. 
We note that the cryptographic techniques used as building-blocks are well studied and fairly standard but are combined in a novel and carefully thought out manner to achieve practical efficiency and meet the goals outlined in \autoref{sec:design}.

 {\renewcommand{\arraystretch}{1.2}%
 \footnotesize
 	\begin{table*}
 		\begin{tabular}{l l  m{8.30cm} }
 			\toprule
 			\textbf{Property} & \textbf{Cryptographic Building Block(s)} & \textbf{Role in \name} \\ [0.5ex] 
 		 	\toprule
 			\arrayrulecolor{bluegray}
 			
 			\textbf{Access Control}
 			
 			 & {Digital signatures \& Encryption} 
 				
 			 &  {Access to data is restricted to the \name protocol by being encrypted and hence inaccessible to researchers directly. Moreover, access to the system is controlled via digital signatures to thwart system abuse by any single researcher in the form of a denial-of-service attack.}\\ 
 			\hline
 			
 			\textbf{Hypothesis tracking} 
 			
 			& \shortstack{Homomorphic Encryption \& Multi-party computation} 
 			
 			& Computation result are revealed via consensus of computing servers, and remain hidden otherwise. Researchers cannot test hypotheses outside of \name because the data remains encrypted and secret at all times.    \\
 			\hline
 			
 			\textbf{Auditability} 
 			
 			& \shortstack{Tamper-proof ledger \& Controlled Decryption} 			
 			& Every statistical test computation is recorded in a publicly readable log at decryption time making test results (and application of false discovery control procedures) auditable by researchers and third parties.  \\
 			\hline
 		\end{tabular}
 		\caption{Summary of cryptographic building blocks and their role in realizing the main properties of~\name.}\label{table:properties}
 	\end{table*}
 }

\subsection{Preliminaries}
\name is constructed using the following building blocks: a tamper-proof ledger for recording all statistical tests executed over a dataset in an auditable fashion, somewhat homomorphic encryption for evaluating statistical tests over encrypted data, and multi-party computation for securely decrypting results and in some cases evaluating parts of the statistical test circuit. \autoref{table:properties} provides a summary of the cryptographic building blocks used and their role in \name. 

\smparagraph{Tamper-proof Ledger} We rely on a tamper-proof ledger abstraction which we call $\Log$. The ledger records statistical test executed on the dataset such that each test result and order of test execution is publicly available and remains immutable once recorded. In \name, $\Log$ can be maintained by a central authority, by the computing servers themselves using a consensus protocol (e.g., Paxos~\cite{lamport2001paxos}), or by independent parties (e.g., using a public Blockchain based on a decentralized consensus protocol such as Nakamoto consensus~\cite{nakamoto2008bitcoin}). Similar to the ledger abstraction used in~\cite{Cecchetti:2017:SCD:3133956.3134010}, our only requirement is that $\Log$ is correct and available.

\smparagraph{Somewhat Homomorphic Encryption}
Somewhat Homomorphic Encryption (SHE) is a special form of asymmetric encryption where a limited number of computations can be performed over the encrypted values without knowledge of the decryption key. In other words, SHE enables the evaluation of a limited class of function on encrypted inputs to obtain an encrypted output. Unlike fully homomorphic encryption~\cite{gentry2009fully}, where arbitrary functions can be evaluated on encrypted inputs (theoretically at least), SHE can only evaluate a handful of multiplication gates but does not incur high computational overhead of fully homomorphic encryption~\cite{sealcrypto}. 
	
\smparagraph{Multi-party Computation}
	Multi-party computation (MPC) is a set of techniques for computing over secret data that is shared amongst a set of parties. The data is encoded and shared in such a way so as to prevent individual parties (or even a subset of parties) from obtaining any information about the (secret) data while still allowing for functions to be evaluated over the encoded data. The importance of MPC is that functions can be evaluated by the parties while ensuring that no party learns the input values. 
 	
	In \name we use the SPDZ~\cite{damgaard2013practical, keller2018overdrive} protocol for evaluating functions in a multi-party setting. SPDZ is the current state-of-the-art when it comes to performing computations and is highly efficient in practice. We provide additional details about the protocol in~\autoref{sec:stattests}.

\subsection{\name protocol} 
\label{subsec:protooutline}
We are now ready to describe the construction of \name. 
The functionality of the \name can be split into three distinct phases: {\it Setup}, \textit{ComputeTest}, and \textit{PerformAudit}.

\smparagraph{Setup} To initialize the system, the data owner encrypts a dataset~$\denc{\Ds}{}$ and shares partial decryption keys with the set of computing servers $\serverset$ such that collectively they may decrypt computation results. 
The data owner then initializes the public log $\Log$ and posts to it a signed message $(0, \bot, \bot)$ that corresponds to the counter of the tests to be executed on $\Ds$, set to zero. In addition to releasing the encrypted dataset, the data owner releases metadata pertaining to $\Ds$ deemed sufficient for researchers to form hypotheses on the dataset (e.g., attribute metadata, etc). We formalize the metadata requirements in~\autoref{sec:security}.

\smparagraph{ComputeTest} 
A researcher uses the encrypted data to evaluate a statistical test~$\codeid$ evaluated over a subset of attributes in $\Ds$. Once the researcher obtains the encrypted result, she sends it to the servers $\serverset$ which engage in the multi-party protocol to decrypt the result. In doing so, the servers make the result publicly available on $\Log$. Recall that the test result cannot be recovered by \emph{any} of the parties individually and requires all servers to reveal their shares in order to recover the result. This specification ensures that each test result is made publicly available to all the servers and researchers and the end of the computation. 
The decrypted result, in conjunction with the existing data written to $\Log$, is the test certificate consisting of the tuple $\cert = (\logid, \codeid, \result)$. 

\smparagraph{PerformAudit}
A test with certificate $\cert = (\logid, \codeid,\result)$ can be audited for
correctness by any entity with access to $\Log$. 
An auditor retrieves all certificates up to $\logid$ from $\Log$ and uses this information to ensure the \ainv procedure was applied correctly when claiming a result is statistically significant. Note that auditing the statistical test results can be done entirely ``offline'' using only information recorded on $\Log$ -- this makes the auditing procedure non-interactive. 

\begin{algorithm}
	\SetAlgorithmName{Procedure}{protocol}{}
	\caption{\texttt{PerformAudit($\cert_\logid = (\logid, \codeid, \result)$)}}
	\label{\nrow}

	$W(\logid-1) \gets $ wealth remaining after applying \texttt{InvestmentRule($\alpha, \beta, \gamma$)} up to the p-value of the $\logid$th test\; 
	\BlankLine
	$\alpha_\logid \gets \min\left(\alpha, \frac{W(\logid-1)\left(1-\beta\right)}{1 + W(\logid-1)\left(1-\beta\right)}\right)$\tcp*{see \autoref{subsec:mcpcontrol}}
     Accept $\cert$ as \emph{valid} if and only if $p_\logid < \alpha_\logid$\;
        
\end{algorithm}


\section{Statistical Tests in \name}
\label{sec:stattests}
In this section we describe the statistical tests we implement in \name and their usage in hypothesis testing. 
We describe four very common statistical tests: \ttest, \pearson test, the \chisq test, and \ftest. While not exhaustive, this set of statistical tests forms a basis for quantitative analysis and covers many cases: from reasoning about population means to analyzing differences between sets of categorical data. These four tests are also widely used in practice and accounted for more than 70\% of all clinical trials from a pool of 1,828 medical studies from a variety of journals~\cite{du2010choosing}. 

 With our selection of tests we  aim to demonstrate the versatility and practicality of \name to real world applications. Our selection captures both tests used on \emph{continuous} and \emph{categorical} data. Many other statistical tests are close variants of the statistical tests we describe in this paper and can be implemented in \name as well.  

\subsection{Dataset Characteristics and Notation}
\label{subsec:datasetproperties}
Before diving into the equations, we first describe how the data is structured in \name, how computations are performed over encoded data, and introduces some additional notation.

\smparagraph{Metadata requirements}
First, we formalize which information about the data is revealed (resp. hidden) from researchers at setup time. 
To allow researchers to form hypotheses on~$\denc{\Ds}{}$, the data owner is required to releases \emph{attribute metadata} about the dataset. This includes, for example, the number of attributes,
the type and domain size, independence from other attributes, etc.). This is to help researchers determine what test to use and what hypothesis to test for. We require that the metadata be such that it is possible to 1) define a hypothesis on $\Ds$ and 2) that it contain the number of rows and columns in the dataset, denoted by $\nrow$, $m$. Moreover, since some tests only work on attributes that are independent of each other, we require the independence of attributes (if any) to be contained therein. Finally, we note that we implicitly set a bound on the attribute domain size (e.g., assume each value can be represented by a 32 bit integer). This is necessary for the correct selection of the parameters at system setup time.

\smparagraph{Computing p-values from test statistics}
We make an important observation which enables us to more efficiently apply FDR control procedures in \name. Each statistical test produces a \emph{test statistic} from which p-values are derived using \emph{degrees of freedom} (a function of the dataset size). Since the p-value can be computed ``in the clear'' using the revealed test statistic, we do not need to compute the p-value using secure computation which drastically reduces the complexity overhead (since obtaining p-values requires lookup tables in practice; an expensive operation to do over encrypted data). We therefore focus on describing the computation of only the test statistic and leave the mapping from test result to an exact p-value implicit as it can be trivially computed.  We note that in certain situations revealing the test statistic (and as a result the p-value) can be too much information since researchers may be able to reconstruct the dataset after observing a sufficient number of queries. We therefore also describe how \name can be adapted to only reveal \emph{statistical significance} which bypasses p-value computations altogether. 

\subsection{\name for the holdout dataset}
While we describe \name as a general scheme for evaluating statistical tests on a dataset, in practice, we envision the dataset $\Ds$ to be split into an exploration dataset $\Ds_{exp}$ and a validation dataset $\Ds_{val}$. The data owner will encrypt the dataset $\Ds_{val}$ and release the encrypted validation dataset along with the (unencrypted) exploration dataset. This way, researchers can explore different hypotheses using $\Ds_{exp}$ and validate their results through \name using $\denc{\Ds}{{val}}$. Indeed, this is the way we envision \name deployed in real world settings as it mirrors current ``best-practices'', but in a provably certifiable way.

\subsection{Hypothesis Testing}
We are now ready to describe the four statistical tests implemented in \name and their usage in hypothesis testing. We refer the reader to~\cite{bertsekas2002introduction} for a additional details on the statistical tests and their applications.  

{\bigskip\noindent\bf \ttest} is used to compare means of two independent samples. The null hypothesis stipulates that there is no difference between the two distributions~\cite{bertsekas2002introduction}. Let $x$ and $y$ be two independent samples from $\Ds$. Denote the mean of $x$, $y$ as $\bar{x}$, $\bar{y}$, and the standard deviation as $s_x$, $s_y$. The test statistic, $t$, is then computed according to:
\begin{equation}
\label{eq:ttest}
 t \triangleq \frac{\bar{x} - \bar{y}}{s_p\sqrt{\frac{            2}{\nrow}}}
\end{equation}
where $s_p = \sqrt{\frac{(\nrow-1)s_x^2 + (\nrow-1)s_y^2}{2\nrow - 2}}$.

{\bigskip\noindent\bf \pearson test} is used to compare the linear correlation between two continuous and independent variables. The test statistic, $r$, lies in the range $[-1,1]$, where either extreme corresponds to negative (resp. positive) correlation between variables~\cite{bertsekas2002introduction}. 
	Let $x$ and $y$ be two independent continuous samples in $\Ds$.
	Denote the mean of $x$ and $y$ as $\bar{x}$ and $\bar{y}$. Pearson's correlation coefficient is computed as follows:
\begin{equation}
\label{eq:pearson}
	r \triangleq \frac{\sum_{i=1}^{\nrow}(x_i - \bar{x})(y_i - \bar{y})}{\sqrt{\sum_{i=1}^{\nrow}(x_i - \bar{x})^2}\sqrt{\sum_{i=1}^{\nrow}(y_i - \bar{y})^2}}
\end{equation}

{\bigskip\noindent\bf \chisq test} is used to determine whether there exists a significant difference between the expected and observed frequencies in a set of observations from \emph{mutually exclusive} categories. The test evaluates the ``goodness of fit'' between a set of expected values and observed values. 

For a collection of $\nrow$ observations, classified into $\ncat$ mutually exclusive categories, where each observed value is denoted by~$x_i$ for $i=1, 2,\dots, \ncat$, we denote the probability that a value falls into the $i$th category by $\eta_i$ such that $\sum_{i=1}^{\ncat} \eta_i = 1$. The \chisq statistic is given by:
\begin{equation}
\label{eq:chisq}
	\chi^2 \triangleq \sum_{i=1}^{\ncat} \frac{(x_i - e_i)^2}{e_i}
\end{equation}
where $e_i = \nrow\eta_i$. 

{\bigskip\noindent\bf \ftest test} is commonly used to determine the fit of a statistical model to a given dataset.  Let $x_i$ for $i = 0,\dots,\ncat$ be independent samples from $\Ds$. Denote the mean of $x_i$ by $\bar{x_i}$ and the mean across all samples by $\bar{x}$. Denote the $j$th value of the $i$th sample by $x_{i,j}$.  The test statistic, $F$, is then computed according to:  

\begin{equation}
\label{eq:ftest}
	F \triangleq {\sum\limits_{i=1}^{\ncat} \frac{n(\bar{x_i} - \bar{x})^2}{\ncat-1}}~\Big/~ {\sum\limits_{i=1}^{\ncat} \sum\limits_{j=1}^{\nrow}\frac{(x_{i,j} - \bar{x_i})^2}{\nrow - \ncat}}
\end{equation}

\subsection{Statistical Test as an Arithmetic Circuit}
We now turn to describing how the statistical tests are computed in \name. The first obstacle in the way of a practical construction is that somewhat homomorphic encryption is only viable for circuits with low multiplicative depth. With practical implementations of homomorphic encryption, it is only possible to evaluate a few (preferably less than 3) levels of multiplication gates where both inputs are encrypted but supports many additions and multiplications by public constants. Observe that each test statistic follows a pattern: they require only a few multiplication per each term in the summation followed by a single division. This makes it possible to efficiently evaluate the numerator and denominator terms of each statistical test equation using at most 2-3 multiplication operations. However, when it comes to the final division gate (where both the numerator and denominator are encrypted), there is no homomorphic operation to perform it outright. 
Instead, existing methods require a recursive approach (converting the division gate to a series of multiplications) which requires a significant number of consecutive encrypted multiplications and bars the way towards any practical implementation. While theoretically we could use  \emph{fully-homomorphic encryption} to evaluate division gates using an iterative approach, such an approach would significantly blow up the computational complexity (by many orders of magnitude) making the construction impractical for real-world settings.

\begin{figure}[t]
	\centering
	\includegraphics[width=1.00\columnwidth]{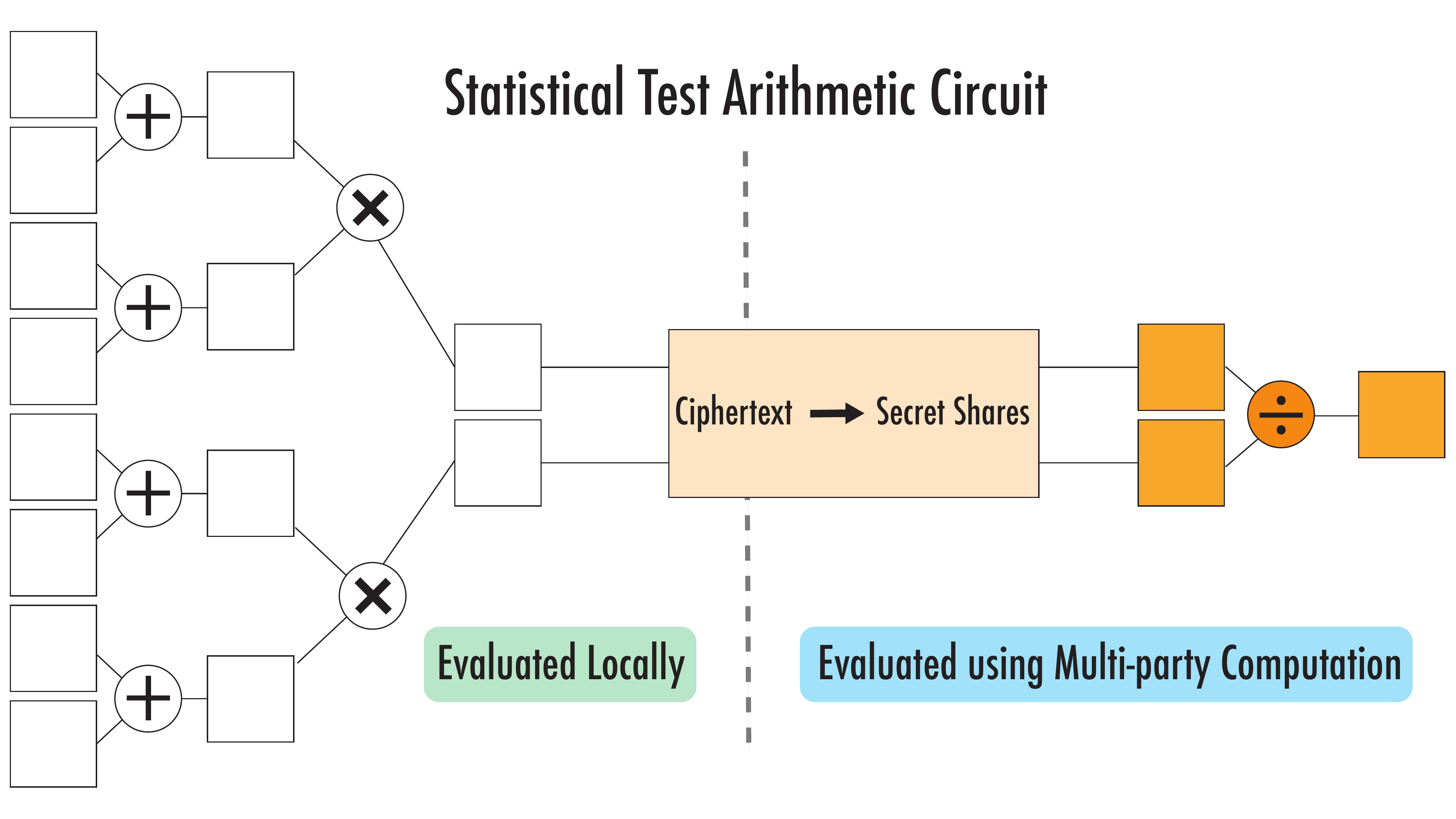}
	\caption{Evaluation of statistical test circuit in \name. Numerator and denominator terms of the test statistic are computed locally by the researcher using the homomorphic properties of the encryption scheme. Computing servers then evaluate the division gate securely using multi-party computation and reveal the result. }
	\label{fig:circuit}
\end{figure}

Instead, we opt for a hybrid approach: we use the decryption parties (parties to whom the shares of the secret key gets distributed to at system setup time) to ``boost'' the computational power of encryption scheme and evaluate \emph{one} division gates. We can use multi-party computation techniques to evaluate division gates in the arithmetic circuit of the statistical test. Since statistical tests that we use to describe \name only require \emph{one division per evaluation}, and at the very end of the computation, we can evaluate the numerator and denominator locally using somewhat-homomorphic encryption and then perform the division using the help of the computing parties.  This leads to a practical solution for evaluating statistical tests precisely because the bulk of the test can be evaluated locally and thus requires minimal additional interaction from the parties (where network latency is a bottleneck). If the majority of the circuit is evaluated locally, the parties are only needed to evaluate division gates and thus only need to receive two ciphertexts (numerator and denominator) which they divide prior to revealing the result. This provides the best of both worlds when it comes to practical efficiency.~\autoref{fig:circuit} illustrates the process of evaluation of a statistical in \name, where the test is represented as an arithmetic circuit. 

\smparagraph{Ciphertext to Share conversion} One caveat, however, is that in order for parties to perform multi-party computation, the inputs (e.g., the numerator and denominator for a devision gate) must be \emph{secret shared} to all the parties. Normally, this is done by a trusted party that splits the plaintext inputs into shares which are then distributed to each party indivually such that any threshold number of parties can pitch-in their shares and collectively recover the secret value. In \name, however, these shares must be created and distributed on-the-fly from the encrypted inputs. To achieve this, we use a multi-party protocol which securely converts an encryption to a set of shares held by parties. This is achieved using a protocol that takes as input the ciphertext encrypted under a public key and results in parties having secret shares of the previously encrypted value. Such protocols have been well-studied and are described in~\cite{cramer2001multiparty}.


\section{Security Analysis}
\label{sec:security}

We now analyze the different security properties outlined in~\autoref{sec:design}. The security of \name hinges on the use of the  somewhat-homorphic ecnryption scheme and the multi-party computation framework used. We opt for the SPDZ multi-party computation framework because it is the current state-of-the-art when it comes to performing efficient MPC computations. We use SPDZ to evaluate the division gates in the statistical test circuit and control the revelation of statistical test results. Security hinges on the fact that given an encrypted dataset, researchers are unable to recover the underlying data to perform covert statistical tests \emph{outside} of \name, i.e., without requesting a decryption from the parties.

\smparagraph{Correctness} 
\name follows the plaintext execution of statistical tests by evaluating an arithmetic circuit of the tests over the encrypted dataset as presented in~\autoref{sec:stattests}. The only difference with a plaintext evaluation compared to the encrypted evaluation is in the accuracy of the results since \name guarantees $O(\fpparam)$-bits of precision which follows directly from the parameters used in instantiating the encryption scheme and SPDZ parties. The precision parameter, $\fpparam$, can be set to provide an arbitrary decimal places of precision. The default is usually set to $2^{40}$ which guarantees at least 12 decimal places of precision and is more than sufficient for p-value calculation in most situations~\footnote{Many p-value calculation tables are rounded to fewer than 5 decimal places~\cite{bertsekas2002introduction}.}. 

\smparagraph{Confidentiality} 
The aim of \name is to enforce p-value calculation in a truthful manner.
It achieves this goal by hiding the content of $\Ds$ and revealing only the metadata on $\Ds$ sufficient to form hypotheses and compute statistical significance. The data owner only needs to reveal the size of the dataset and attribute metadata information on the contents of~$\Ds$ (e.g., characteristics of each attribute, domain size, and independence from other attributes)~\footnote{For example, metadata for an ``age'' attribute may be the set \texttt{\{AttrType: Age, NumericRange: 0-110\}}). We stress, however, that the metadata can be made general and independent of the actual values found in $\Ds$, when deemed of no importance to forming hypotheses.}. 
The metadata is succinctly captured as a function $\lsetup:\Ds \rightarrow \{0,1\}^*$,
that takes as input the dataset and returns the number of rows and attributes $\nrow$, $\ncat$ and other metadata decided on by the data owner. At setup time, this function captures the exact ``leakage'' of $\Ds$ as nothing beyond that is revealed.  

There is additional leakage of information on $\Ds$ that results from the statistical test computation and is captured by the function $\lcomp:(\Ds \times \codeid) \rightarrow \codeid(\Ds)$ which is the result of the statistical test $\codeid$ computed on $\Ds$. 
Confidentiality of the dataset then follows directly from the fact that the only information revealed by computing a statistical test (represented as an arithmetic circuit) over the encrypted dataset $\enc{\Ds}$ is the output $\codeid(\Ds)$ (i.e., the result of the statistical test $\codeid$) and nothing else. If this is deemed too much, the leakage can be reduced to just revealing significance (we explain this in the following subsection) in which case \emph{only a single bit of information is leaked when performing a statistical test} through \name.

\smparagraph{Access Control} Though anyone can perform a computation over $\enc{\Ds}$, the computing servers can enforce ``rate limits'' on researchers or refuse to reveal results to researchers who attempt to abuse the system. This is an important feature to avoid a situations where a malicious researcher exhausts the \ainv procedure's alpha-wealth and prevents other researchers from testing their hypotheses. While describing the exact access control policies that could be enforced is outside the scope of this work, we emphasize that such policies can be useful in real-world situations and can be integrated in the design of \name by default. For example, the system can be easily made to enforce daily quotas on the number of statistical tests that any single researcher can evaluate over a dataset.

\smparagraph{Auditability}
All computations are performed securely over the encrypted dataset $\denc{\Ds}{}$ where we assume at least one of the computing servers follows the protocol honestly. Thus, every resulting decryption is written to $\Log$ which guarantees auditability of all the statistical test result computed on $\Ds$. For every certificate $\cert =(\logid, \codeid, \result)$, the sequence of resulting computations numbered $1$ to $\logid - 1$ can be used to certify the validity of $\cert$ by auditing the application of the \ainv procedure as shown in \autoref{subsec:mcpcontrol}. It is important to note that auditing does not require interaction from the computing parties and can be done fully ``offline'',- using only publicly available information.

\subsection{1-bit computation leakage}
In some situations, leaking the result of the statistical test may be deemed to reveal too much about the dataset $\Ds$. For example, there may be situations where researchers can exploit the information revealed by a statistical test to make an estimated guess about the significance of a yet-not-computed statistical test (effectively cheating the control procedure). If this is deemed a risk, then the computation leakage can be reduced to $\lcomp:(\Ds \times \codeid) \rightarrow \{0,1\}$ where \emph{the only output of a computation through \name is a single bit denoting statistical significance and nothing else.} This can be achieved by performing a comparison of the encrypted result $\result$ with a significance threshold $t$ using secure computation (right after evaluating the division gate, see~\autoref{fig:circuit}) and only revealing the result of this comparison. This potentially sacrifices the utility of the result since nothing beyond statistical significance is revealed but provides the minimum leakage possible (a single bit). We note that the \ainv procedure still works (and can be audited as before) because the procedure only requires the result of this threshold comparison, and not the p-value itself.


\begin{figure*}[t]
	\centering
	\includegraphics[width=1.00\textwidth]{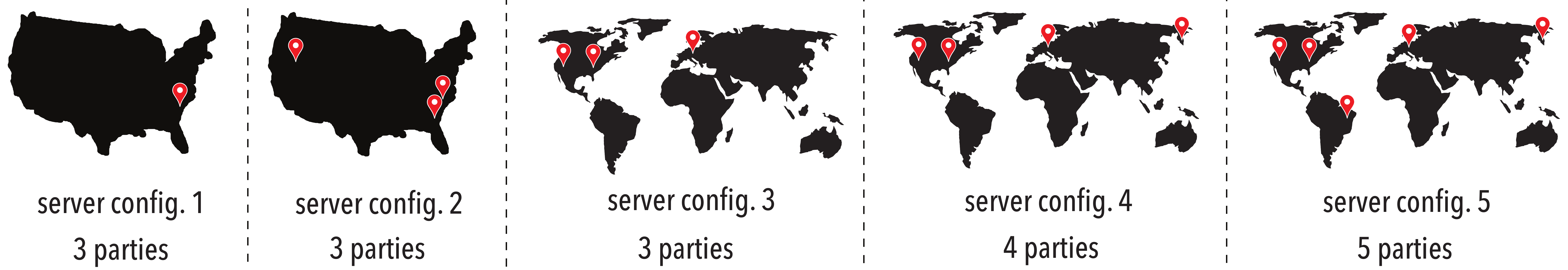}
	\caption{Server configurations used in the evaluation of \name.}
	\label{fig:config}
\end{figure*}

\section{Experimental evaluation}
\label{sec:experiments}

We now turn to demonstrating the applicability of \name as a system for providing auditable results to statistical tests in an efficient way. Our goal is to show that \name has minimal overhead on the data owner and researchers. Moreover, since we envision a deployment of \name where the computing servers are far apart (e.g., universities in the United States and in Europe), network latency has the potential to incur significant performance overheads which could be a problem in practice and we must therefore evaluate carefully. Finally, we wish to evaluate the effectiveness of the \ainv procedure for controlling the FDR to illustrate why this procedure is an ideal pairing for \name. 

\smallskip
We outline the goals of our evaluation as follows:
\begin{itemize}
	\item Thoroughly evaluate \name on a variety of datasets to measure the overall runtime for all three statistical tests. Specifically, we wish to show that \name can perform in realistic deployment scenarios where the datasets are sufficiently large in size and servers used in the final steps of the computation are far apart from one another. 
	\item Illustrate the effectiveness of the \ainv procedure when it comes to controlling false discoveries under different dataset characteristics to show why \name would be effective at preventing p-hacking when used as described. 
	\item Measure the impact that network latency, number of computing servers, and other parameters, introduce and the effects on the practicality of the \name.
	\item Measure the overhead imposed on the data owner at system setup and on researchers during the evaluation of statistical tests.
\end{itemize}

\subsection{Implementation and Environment}
We implement \name in C++ using the open source homomorphic encryption library SEAL (v3.3.0)~\cite{sealcrypto} and SPDZ MPC engine (\url{https://github.com/KULeuven-COSIC/SCALE-MAMBA}). The implementation of statistical tests computes the equations described in~\autoref{sec:stattests}: the SEAL library is used by the data owner to encrypt the data, researchers use the encrypted dataset to evaluate statistical tests, and the final result is computed and revealed by parties running the SPDZ protocol. 
SEAL does not currently support threshold decryption as described in~\cite{jain2017threshold, boneh2018threshold}, we therefore simulate the evaluation of the share-conversion protocol (see~\autoref{fig:circuit}) using the SPDZ parties and a single decryption key by emulating the ciphertext to share conversion protocol run by the computing parties. 

The homomorphic evaluation of the statistical tests were conducted on a single machine with Intel Xeon E5 (Sandy Bridge) @ 2.60GHz processor (30 cores per machine) and 120GB of RAM, running Ubuntu 18.04 LTS. The cost of running the machine was estimated at \$914.98/month.

Each computation party was instantiated on a machine with Intel Xeon E5 (Sandy Bridge) @ 2.60GHz processors (8 cores per machine) and 15GB of RAM, running Ubuntu 18.04 LTS. The cost of running each party was estimated at \$109.79/month.

Unless otherwise stated, each result is the average over five separate trial runs (for each combination of parameters).
 
\subsection{Datasets}
We evaluate each statistical test on synthetic and real-world datasets. However, we note that \emph{runtime} performance is not impacted by the distribution of the underlying data. Since the data itself is encrypted, all computations run in the same amount of time because the evaluation of the statistical tests is done over the ciphertexts. 
We evaluate \name on one real world dataset which we make the same size as one of the synthetic datasetsd to illustrate this fact. The characteristics of the datasets used in our evaluation are summarized in Table~\ref{table:datasets}.

The real world dataset was obtained from the UC Irvine Machine Learning Repository~\cite{Asuncion+Newman:2007}. For experiments involving \ttest, \pearson test, and \ftest (which require \emph{continuous} data), we use the Abalone dataset consisting of a total of nine continuous measurements and $4,177$ rows.  We truncate the dataset down to $1,000$ rows to provide a comparison baseline with our synthetic dataset cont\_1k. We do this to illustrate the fact that the only factor influencing the computational overhead is the size of the dataset --- the synthetic datasets do not advantage \name in any way. 

\smparagraph{Synthetic datasets}
We generate three continuous synthetic datasets containing 1,000, 5,000 and 10,000 rows, respectively, with random real value entries sampled from a normal distribution. We use these datasets to evaluate \ttest, \pearson test, and \ftest. 
For evaluating \chisq, we generate three categorical datasets containing $8$ mutually exclusive categories. We evaluate the \chisq test on a subset of $2$, $5$, and $8$ mutually exclusive categories to demonstrate the impact of varying the number of selected attributes over which the test is performed.

\begin{table}[h]
	\centering
	\begin{tabular}{lllll}
		\toprule
		{\bf Dataset}  & {\bf\# rows} &  {\bf\# attrs.}  & {\bf Data type} \\
		\toprule
		abalone	        & 1000$^*$ &  8   & continuous  \\
		cat\_1k      & 1000  &  8  & categorical \\
		cat\_5k 	    & 5000 	&  8  & categorical \\
		cat\_10k     & 10000 &  8  & categorical	 \\
		cont\_1k     & 1000  &  3   & continuous  \\
		cont\_5k 	& 5000 	&  3   & continuous  \\
		cont\_10k 	& 10000 &  3   & continuous  \\
		\bottomrule
	\end{tabular}
	\caption{Summary of dataset characteristics.}
	\label{table:datasets}
	\small{$^*$ We truncate this dataset to 1000 rows to illustrate the fact that computation time is entirely independent of the data distribution (as it should be given that the data is encrypted)}
\end{table}

\subsection{Experiment Setup}
To ensure our experiments emulate a potential deployment scenario (e.g., where servers are maintained by universities in different geographic regions), we run our experiments with five different configurations. We set up servers in three different locations in the USA (South Carolina, Virginia, Oregon) as well as Germany, Brazil, and Japan with 3-5 parties depending on the configuration. We report each of the five configurations in Figure~\ref{fig:config}. Our choice of server configurations allows us to evaluate the effects of network latency in a deployed scenario. The first configuration consists of all three servers located in the same datacenter (South Carolina) which results in parties being connected on the local area network (LAN) with ``ideal'' network latency. We use this configuration as as reference point to analyze the impact of settings where servers are further apart. 

\subsection{Parameters}
To ensure all tests are evaluated in a way that enables comparisons between results, we fix the parameters ahead of time to satisfy the requirements imposed by all four statistical tests and datasets.  Specifically, we set fixed-point scaling factor $\fpparam = 40$ which provides up to 12 decimal places of precision for each computation and is sufficient for computing precise p-values.
We set the statistical security parameter $\ssparam = 40$ which provides $40$-bits of statistical security during multi-party computations and is the default choice~\cite{damgaard2013practical}. In words, this parameter ensures that the probability of a computation leaking information about a secret value is less than $\frac{1}{2^{\ssparam}}$, which is negligible in $\ssparam$.

\begin{figure}
\centering
  \includegraphics[width=\linewidth]{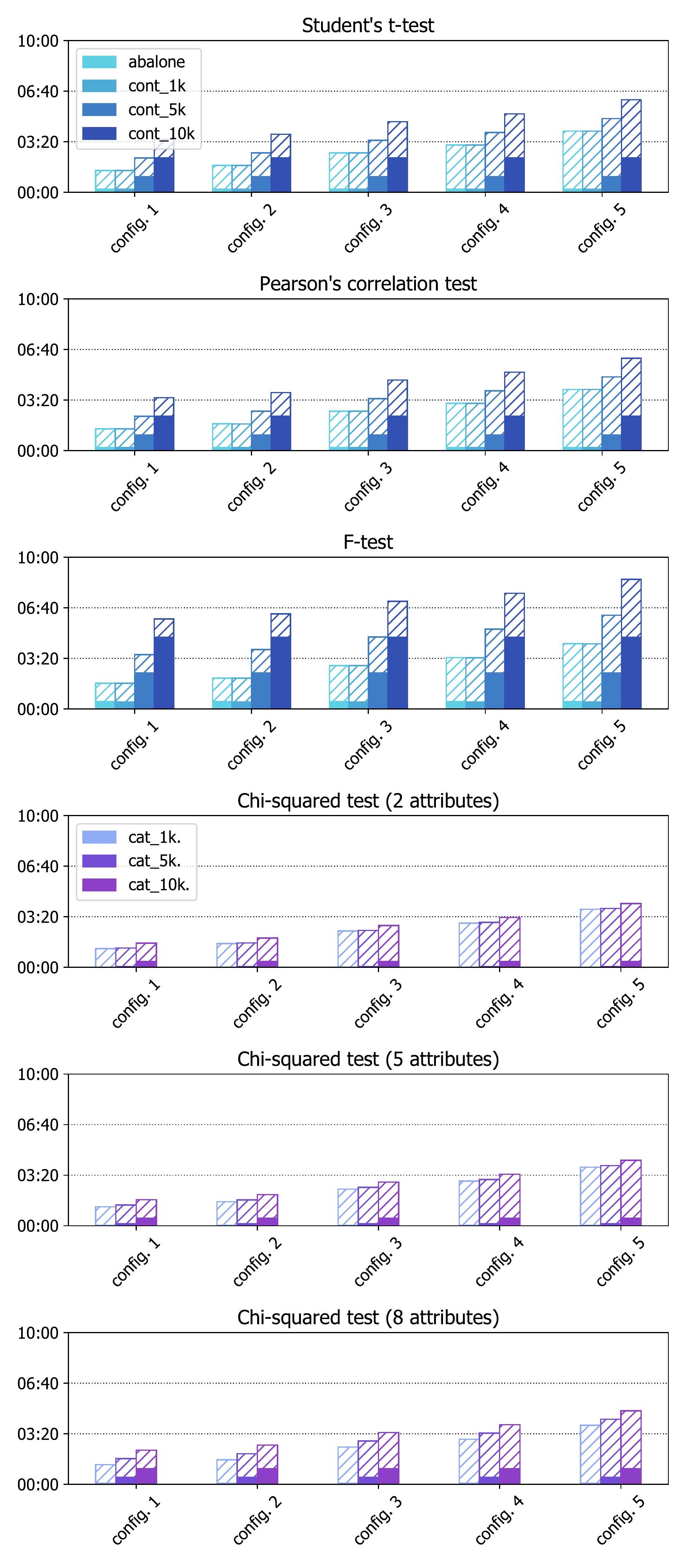}
  \caption{Total computation time (mm:ss) required for each test as a function of the computing server configuration and dataset selection. Offline runtime is presented in solid color; online runtime is cross-hatched. }
  \label{fig:offlinetesttime}
\end{figure}

\begin{figure}
\centering
  \includegraphics[width=\linewidth]{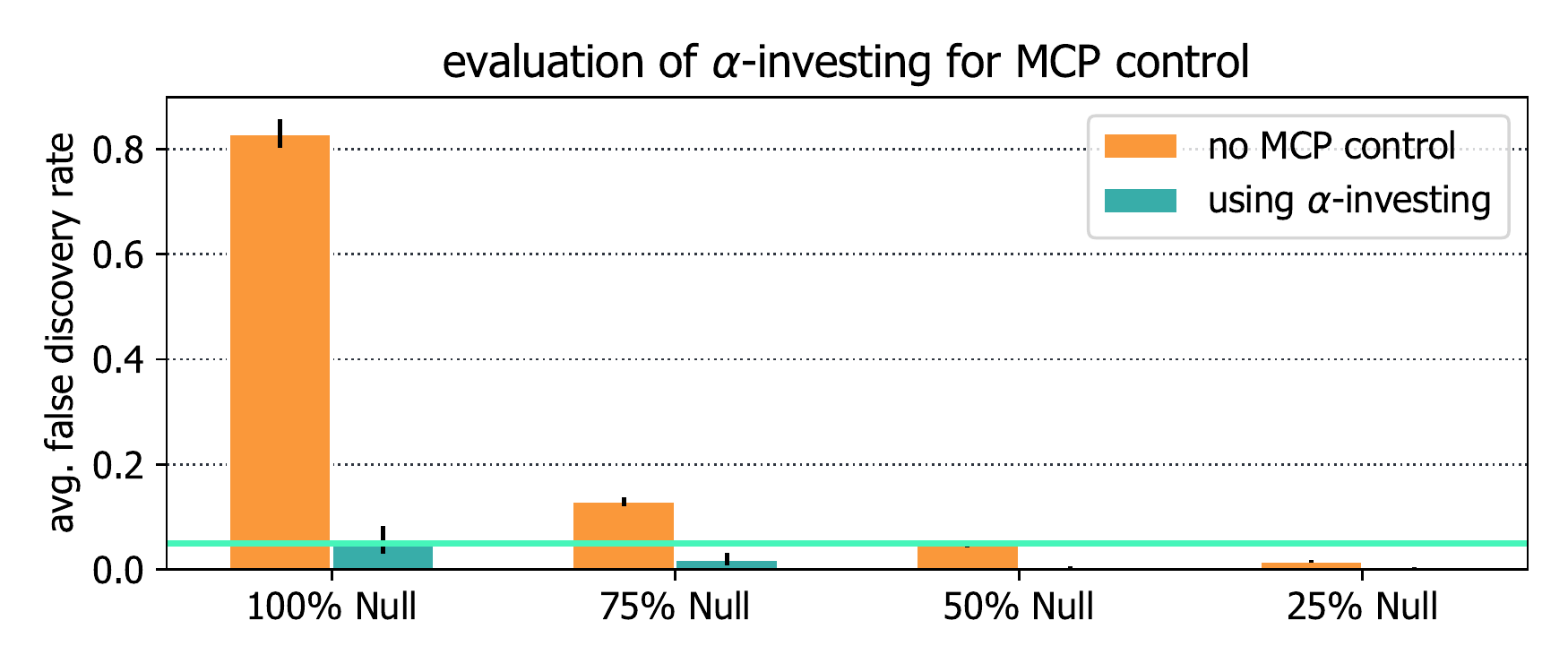}
  \caption{Comparison of average FDR with and without the \ainv control procedure on synthetic datasets with varying percentage of null hypotheses. Green line indicates the $\bold{\alpha = 0.05}$ threshold. }
  \label{fig:alphainvst}
\end{figure}

\begin{figure}
\centering
  \includegraphics[width=\linewidth]{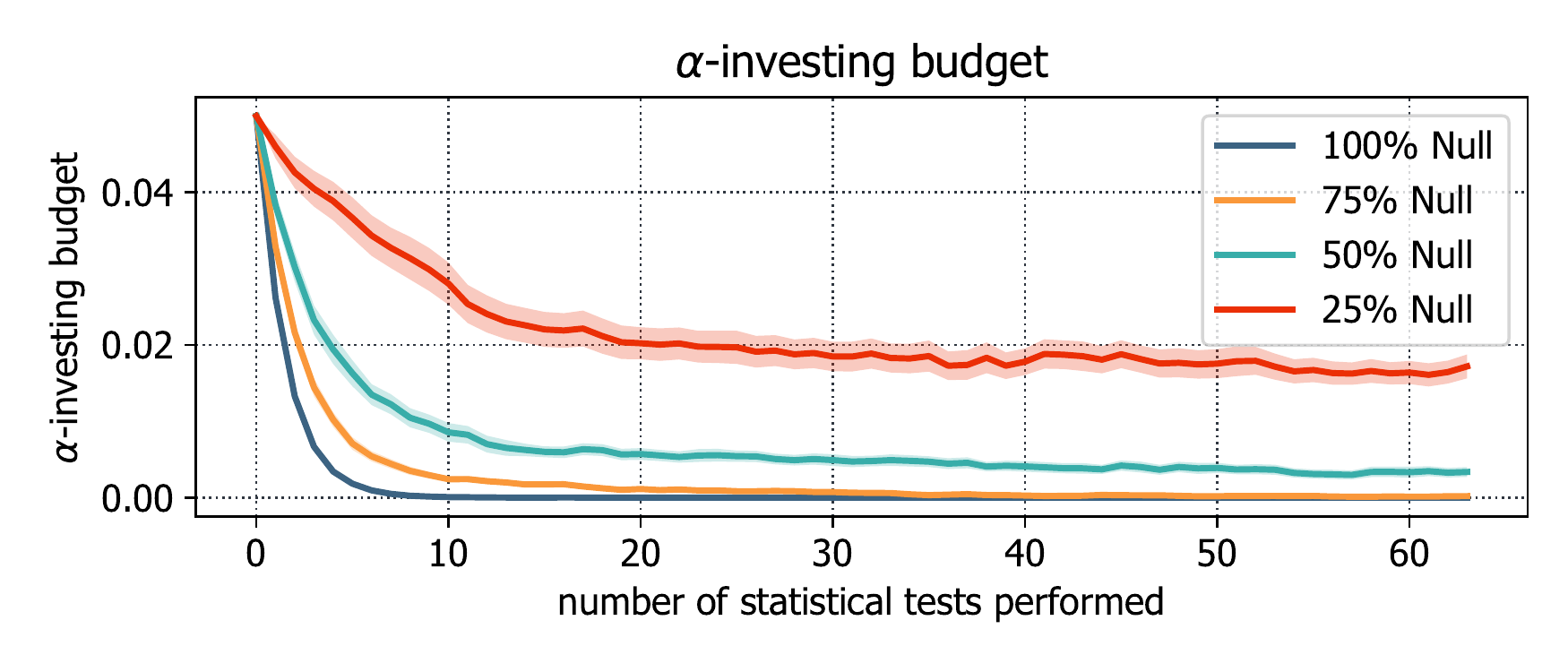}
  \caption{Remaining \ainv budget as a function of the number of tests computed over synthetic datasets with varying percentage of null hypotheses.}
  \label{fig:alphabudget}
\end{figure}

\subsection{Results}

We describe the results in four parts which roughly correspond to the goals outlined at the beginning of this section. 
 
\begin{figure}
\centering
  \includegraphics[width=\linewidth]{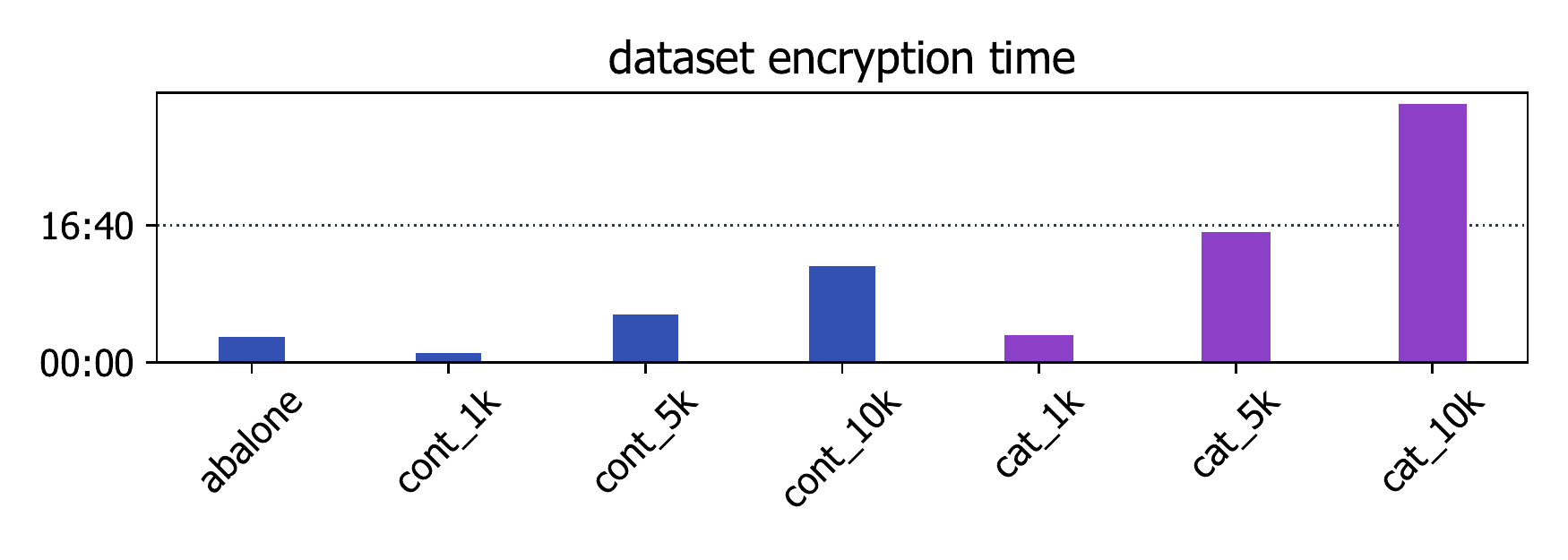}
  \caption{Data owner setup time (mm:ss) for each dataset used in the evaluation. }
  \label{fig:setuptime}
\end{figure}

\smparagraph{System setup}
Across all experiments, the setup time \name (i.e., time required to encode and encrypt the data) remained below $32$ minutes per dataset. We report the setup time for each dataset in~\autoref{fig:setuptime}. We stress that this is only a one-time operation needed at setup time: after the dataset is encrypted, the data owner goes offline and does not need to interact with researchers or perform any further computation.  

\smparagraph{Computation runtime}
Since we implement \name using the SEAL homomorphic encryption library and the SPDZ multi-party computation engine, we split the runtime evaluation into two phases: \emph{offline} and \emph{online}. The offline phase is where researchers evaluate the statistical test arithmetic circuit using homomorphic encryption to obtain the numerator and denominator terms. In the online phase, the researchers provide the evaluated numerator and denominator ciphertexts to the computing parties which then engage in a multi-party computation to evaluate the division and reveal the statistical test result. We report the runtime of each statistical test in~\autoref{fig:offlinetesttime}. 

The runtime of computing statistical tests ranged from a few seconds to approximately 9 minutes on the larger datasets. Even though this is several orders of magnitude overhead compared to computing a statistical test directly over the \emph{unencrypted} data (e.g., using SciPy), it is not unreasonably high when considering practical circumstances. Researchers in the real world are likely to only evaluate a few statistical tests on a dataset per day which \name is adept at handling, even with many researchers using the system. The results also illustrate the overhead of performing multi-party computation. In~\autoref{fig:offlinetesttime} we see that the online time accounted for a significant fraction of the computation time, even though it is only used for evaluating the division gate and revealing the result. This confirms that evaluating the full statistical test over shared data using multi-party computation would incur an unreasonably high overhead on the runtime for a practical application.

\smparagraph{Discovery certification}
To demonstrate how \name performs at its main goal of discovery certification, we evaluate the \ainv procedure for controlling false discoveries. For evaluating the FDR with and without the \ainv control procedure, we generate one hundred synthetic datasets with values ranging between 0 and 100 across 64 attributes and 1,000 rows. For each dataset we perform 64 ``multiple comparisons'' using each of the three tests over a subset of 2 attributes.  
We repeat this process for four dataset collections where each collection contains a different percentage of true null hypotheses (e.g., $100\%$ implies completely random data and so all ``discoveries'' are false). We report the results in Figure~\ref{fig:alphainvst}. The results show that the \ainv procedure indeed bounds the expected FDR to the specified alpha ($\alpha=0.05$), even when all discoveries in the data are false. 
We additionally report the $\alpha$-budget remaining after each test (following Procedure~\ref{alg:investmentrule}) in Figure~\ref{fig:alphabudget}. We use the same generated datasets and set the initial wealth $W(0) = 0.05$ and parameters $\beta = 0.5$, $\gamma = 0.0125$. As expected, the budget is never fully depleted but quickly decreases in the case of datasets with a large portion of null hypotheses.


\section{Related Work}
Much of the related work surrounding \name either focuses on private computations using 
homomorphic encryption or non-cryptographic methods for preventing p-hacking such as pre-registration of hypotheses. To our knowledge, 
there has been no prior work using cryptographic techniques for certifying the validity of statistical tests in a provable and fully auditable way. 

Sornette~\textit{et al.}~\cite{sornette2009financial} describe an approach for demonstrating the validity of stock market bubble predictions \emph{a posteriori}. Their method relies on hashing documents containing financial forecasts which get released after the forecasted date has passed. Coupled with a trusted entity for storing the hashes, their method provides a means of validating the historical accuracy of predictions in an unbiased way. While their solution isn't directly applicable to hypotheses testing, the main goal of their work is comparable to ours. 

Dwork~\textit{et al.}~\cite{dwork2015preserving, dwork2017guilt} propose a method that
constrains analyst's access to a hold-out dataset as follows.
A trusted party keeps a hold-out dataset and answers
up to $m$ hypotheses using a differentially private algorithm.
Here, $m$ depends on the size of the hold-out data and
the generalization error that one is willing to tolerate.
Hence, compared to our decentralized approach,
this method assumes \emph{trust into a single party} that (1) does not release the dataset to the researchers and (2) stops answering queries when more than $m$ hypotheses queries.
Moreover, the method introduces random noise into the results which can potentially alter the outcome of the computed test.

Lauter~\textit{et al.}~\cite{lauter2014private}
and Zhang~\textit{et al.}\cite{zhang2015} propose
the use of homomorphic encryption to compute statistical tests on encrypted
genomic data. However, the constructions are not intended for validating statistical 
testing procedures but rather for guarding the privacy of patient data.
Furthermore, Lauter~\textit{et al.}~make several simplifications such as not
performing encrypted division (rather performing arithmetic division in the clear), imposing assumptions on the data, etc., making their solution less general
compared to methods for computing statistical tests in \name. 

Homomorphic encryption and multi-party computation techniques have been used for outsourcing machine learning tasks of private data~\cite{wu2012using,bost2015machine,cryptonets,Graepel2013}.
Our work, however, crucially relies on the decryption
functionality being decentralized in order to control and keep track of data exposure.

Other related work surrounds \emph{non-cryptographic} techniques to guard against false discoveries in statistical analysis. This includes procedures such as the \ainv procedure~\cite{zhao2017controlling,foster2008alpha} and other methods for bounding the number of false discoveries~\cite{dunn1961multiple, benjamini1995controlling}. When applied properly these techniques prevent p-hacking in an idealized scenario but they provide no guarantee that the methods were are applied correctly which is why the system we describe becomes a necessity. 

Finally, we note that preserving privacy of dataset values when releasing results
of statistical tests~\cite{pmlr-v48-rogers16,Johnson:2013:PDE:2487575.2487687} is orthogonal to our work.


\section{Conclusions}

We present \name, a system that certifies hypothesis testing using cryptographic techniques and a decentralized certifying authority to eliminate avenues for p-hacking and other forms of data dredging.
\name computes statistical test over encrypted data by combining somewhat homomorphic encryption with multi-party computation in a way that enables efficient and practical evaluation yet provides full auditability of results. We show that \name supports a wide range of statistical tests used in practice for hypothesis testing and evaluate four of the most common tests used by researchers. Our experiments demonstrate the feasibility of such a system for real-word settings and only incurs a small computational overhead on researchers evaluating tests. As such, we believe that \name is a viable solution to prevent p-hacking in a \emph{provable} way, promoting accountability and reproducibility in scientific studies. With \name, data can be released for public use with the guarantee that the insights extracted from the data are correct and all necessary false discovery control procedures were applied. \name is the first system which address p-hacking and fosters reproducibility by providing a certificate-of-correctness and enabling \emph{post hoc} auditing of all actions taken by researchers.  

\begin{center}
{$\bigstar$} 	
\end{center}



\balance

\bibliographystyle{abbrv}
\bibliography{bibtexs/relatedwork,bibtexs/phackingproblem,bibtexs/stattests,bibtexs/blockchains,bibtexs/mpc}
 
\end{document}